\documentclass[epj,twocolumn]{webofc}
\usepackage[varg]{txfonts}   
\woctitle{Time machines and traversable wormholes in modified theories of gravity}
\begin{document}
\title{Time machines and traversable wormholes in modified theories of gravity}
%
%

\author{Francisco S.N. Lobo\inst{1}\fnsep\thanks{\email{flobo@cii.fc.ul.pt}}
}
\institute{Centro de Astronomia e Astrof\'{\i}sica da
Universidade de Lisboa, \\Campo Grande, Ed. C8 1749-016 Lisboa,
Portugal}

\abstract{

We review recent work on wormhole geometries in the context of modified theories of gravity, in particular, in $f(R)$ gravity and with a nonminimal curvature-matter coupling, and in the recently proposed hybrid metric-Palatini theory. In principle, the normal matter threading the throat can be shown to satisfy the energy conditions and it is the higher order curvatures terms that sustain these 
wormhole geometries. 
We also briefly review the conversion of wormholes into time-machines, explore several of the time travel paradoxes and possible remedies to these intriguing side-effects in wormhole physics.

}
\maketitle
\section{Introduction}\label{intro}

Much interest has been aroused in traversable wormholes since the seminal article by Morris and Thorne \cite{Morris:1988cz}. These solutions are multiply-connected and probably involve a topology change, which by itself is a problematic issue \cite{DeBenedictis:2008qm}. However, one of the most fascinating aspects of wormholes is their apparent
ease in generating closed timelike curves (CTC) \cite{mty}. A CTC allows time travel, in the sense that an observer that travels on a trajectory in spacetime along this curve, may return to an event before his departure. This fact apparently violates causality, so that time travel and the associated paradoxes have to be treated with great caution.

Another interesting feature of traversable wormholes is that they are supported by {\it exotic matter} \cite{Morris:1988cz}. The latter is defined as matter that violates the null energy condition (NEC), which involves a stress-energy tensor such that $T_{\mu\nu}k^\mu k^\nu <0$, where $k^\mu$ is any null vector. In fact, as the violation of the energy conditions is usually considered a problematic issue in the literature, it is useful to minimize its usage. To this effect, one may consider stationary axisymmetric solutions \cite{Teo:1998dp}, where it was shown that the exotic matter lies in specific regions of spacetime, so that an observer traversing the wormhole may avoid the null energy condition violating regions altogether. It was also shown in the context of evolving wormhole spacetimes, that dynamic wormhole geometries exist for a finite, arbitrarily large, interval of time, with the required stress-energy tensor satisfying the NEC \cite{Kar:1994tz}. One may also minimize the usage of exotic matter in the context of thin-shell wormholes, by using the cut-and-paste procedure \cite{WHthinshell}. In this case, the exotic matter is concentrated at the throat of the wormhole, which is localised on the thin shell. 

An alternative approach lies within modified gravity, where in principle it is possible to impose that normal matter satisfies the energy conditions, and it is the higher order curvature terms that support these exotic spacetimes. In modified gravity, the gravitational field equation may be expressed as $G_{\mu\nu}\equiv R_{\mu\nu}-\frac{1}{2}R\,g_{\mu\nu}= \kappa^2 T^{{\rm eff}}_{\mu\nu}$, where $\kappa^2 =8\pi G$ and $ T^{{\rm eff}}_{\mu\nu}$ is the effective stress-energy tensor, which contains the higher order curvature terms. Therefore, in modified gravity, it is the effective stress-energy tensor that violates the NEC, i.e., $T^{{\rm eff}}_{\mu\nu} k^\mu k^\nu < 0$, and that sustains the wormhole geometry. Recently, this approach has been extensively analysed in $f(R)$ gravity \cite{modgravity1}, curvature-matter couplings \cite{modgravity2}, conformal Weyl gravity \cite{modgravity3}, in braneworlds \cite{modgravity4}, and hybrid metric-Palatini gravity \cite{Capozziello:2012hr}, amongst other contexts.  

In this paper, we review modified gravity scenarios in the context of wormhole geometries. In addition to this, we also briefly review some interesting features related to time travel and its paradoxes in the context of wormhole physics.

\section{Wormholes in modified gravity}

\subsection{Spacetime metric}

Consider the wormhole geometry given by the following static and
spherically symmetric metric
\begin{equation}
ds^2=-e^{2\Phi(r)}dt^2+\frac{dr^2}{1-b(r)/r}+r^2\,(d\theta^2 +\sin
^2{\theta} \, d\phi ^2) \,,
    \label{metric}
\end{equation}
where $\Phi(r)$ and $b(r)$ are arbitrary functions of the radial
coordinate, $r$, denoted as the redshift function, and the shape
function, respectively \cite{Morris:1988cz}. The radial coordinate
$r$ decreases from infinity to a minimum value $r_0$, which represents the wormhole throat, where $b(r_0)=r_0$, and then increases from $r_0$ back to infinity. For the wormhole to be traversable, one must demand that there are no horizons present, so that $\Phi(r)$ must be finite everywhere.

Taking into account embedding diagrams, in order to have a flaring-out of the throat, one verifies that  the following condition is imposed $(b-b^{\prime}r)/b^{2}>0$ \cite{Morris:1988cz}, which reduces to $b^{\prime}(r_{0})<1$ at the throat $r_0$. The latter flaring-out condition, through the Einstein field equation, imposes the NEC violation in classical general relativity. Another condition that needs to be satisfied is $1-b(r)/r>0$.

\subsection{Wormhole geometries in $f(R)$ gravity}

Recently, a renaissance of modified theories of gravity has been verified in an attempt to explain the late-time accelerated expansion of the Universe (see Ref. \cite{Lobo:2008sg} and references therein for more details). In particular, $f(R)$ gravity has been extensively analysed, as it possesses appealing features in that it combines mathematical simplicity and a fair amount of generality. 

More specifically, $f(R)$ gravity consists of generalizing the Hilbert-Einstein gravitational Lagrangian density and is given by the following action:
\begin{equation}
S=\frac{1}{2\kappa^2}\int d^4x\sqrt{-g}\;f(R)+\int d^4x\sqrt{-g}\;{\cal L}_m(g_{\mu\nu},\psi)
\,,
\end{equation}
where $f(R)$ is an arbitrary function of the scalar invariant; ${\cal L}_m$ is the matter Lagrangian density, in which matter is minimally coupled to the metric $g_{\mu\nu}$ and $\psi$ collectively denotes the matter fields.

The gravitational field equations, which are obtained by varying the action with respect to the metric, are given by the effective Einstein field equation $G_{\mu\nu}= T^{{\rm eff}}_{\mu\nu}$, where we have considered $\kappa^2=1$ for simplicity, and the effective stress-energy tensor, $T^{{\rm eff}}_{\mu\nu}$, is defined as
\begin{equation}
T^{{\rm eff}}_{\mu\nu}= \frac{1}{F}\left[T^{(m)}_{\mu\nu}+\nabla_\mu \nabla_\nu F
-\frac{1}{4}g_{\mu\nu}\left(RF+\nabla^\alpha \nabla_\alpha F+T\right) \right]    \,,
    \label{gravfluid}
\end{equation}
where $F=df/dR$.

Consider that the redshift function is constant, $\Phi'=0$, which simplifies the
calculations considerably. Note, that if $\Phi'\neq 0$, the field equations become forth
order differential equations, and become quite intractable.
Thus, the gravitational field equations can be rewritten in the following manner:
\begin{eqnarray}
\label{generic1} \rho(r)&=&\frac{Fb'}{r^2}\,,
       \\
\label{generic2}
p_r(r)&=&-\frac{bF}{r^3}+\frac{F'}{2r^2}(b'r-b)-F''\left(1-\frac{b}{r}\right)
     \,,   \\
\label{generic3}
p_t(r)&=&-\frac{F'}{r}\left(1-\frac{b}{r}\right)+\frac{F}{2r^3}(b-b'r)\,,
\end{eqnarray}
where $\rho(r)$ is the energy density, $p_r(r)$ and $p_t(r)$ are the radial and tangential pressures, respectively. The field equations (\ref{generic1})-(\ref{generic3}) provide the generic distribution of matter threading the wormhole, as a function of the shape function and the specific
form of $F(r)$. Thus, by specifying the above functions, one deduces the matter content of the wormhole.

The violation of the NEC in $f(R)$ gravity imposes that $T^{{\rm eff}}_{\mu\nu} \, k^\mu k^\nu< 0$, which yields the generic condition:
\begin{eqnarray}
\frac{1}{F}\left(T^{(m)}_{\mu\nu}\, k^\mu k^\nu +\, k^\mu k^\nu \nabla_\mu \nabla_\nu F 
\right) < 0   \,.
    \label{NECgravfluid}
\end{eqnarray}
Note that in general relativity, i.e., $f(R)=R$, we regain the condition for the matter stress-energy tensor violation of the NEC, i.e., $T^{(m)}_{\mu\nu} \, k^\mu k^\nu< 0$. 
The condition (\ref{NECgravfluid}) has been extensively explored~\cite{modgravity1}, and specific wormhole solutions have been found. In principle, one may impose the condition $T^{(m)}_{\mu\nu} k^\mu k^\nu\ge 0$ for the normal matter threading the wormhole. 

Thus, in the context of $f(R)$ modified theories of gravity, one may in principle impose that the normal matter threading the wormhole, given by equations (\ref{generic1})-(\ref{generic3}), satisfy the energy conditions, and in particular, the NEC given by
\begin{equation}
\rho+p_r= \frac{1}{2r^3}\left (2F+rF' \right)(b'r-b)+F'' \left( 1-\frac{b}{r} \right)  > 0   \,.
    \label{NECfluid}
\end{equation}
Thus, it is the higher order curvature terms, interpreted as a gravitational fluid, that sustain these non-standard wormhole geometries. We refer the reader to \cite{modgravity1} for more details.

\subsection{Nonminimal curvature-matter coupled wormholes}

Motivated by the dark matter and dark energy problems facing modern cosmology, a generalization of 
$f(R)$ gravity has recently been proposed, involving a nonminimal curvature-matter coupling 
\cite{Bertolami:2007gv}. The action is given  by
\begin{equation}
S=\int \left\{\frac{1}{2}f_1(R)+\left[1+\lambda f_2(R)\right]{\cal
L}_{m}\right\} \sqrt{-g}\;d^{4}x~,
\end{equation}
where $f_i(R)$ (with $i=1,2$) are arbitrary functions of the Ricci
scalar $R$; the coupling constant $\lambda$ characterizes the strength of the interaction between $f_2(R)$ and the matter Lagrangian.

For simplicity, consider the specific case of $f_1(R)=f_2(R)=R$, and the Lagrangian form of ${\cal L}_m=-\rho(r)$ \cite{Bertolami:2008ab}. As before, the gravitational field equation can be expressed as an effective Einstein field equation $G_{\mu\nu}\equiv R_{\mu\nu}-\frac{1}{2}R\,g_{\mu\nu}= T^{{\rm eff}}_{\mu\nu}$, where the effective stress-energy tensor is given by
\begin{equation}
T^{{\rm eff}}_{\mu\nu}= (1+\lambda R)T_{\mu \nu }^{(m)}+ 2\lambda \left[\rho R_{\mu\nu}-(\nabla_\mu \nabla_\nu-g_{\mu\nu}\square)\rho\right]\,.
    \label{efffield2}
\end{equation}
Here, we consider that the redshift function is  $\Phi'=0$, for simplicity, so that the curvature scalar, $R$, for the wormhole metric (\ref{metric}) reduces to $R=2b'/r^2$. The case of $\Phi' \neq 0$ has also been recently analysed \cite{Bertolami:2012fz}.

Thus, the gravitational field equations provide the following expressions
\begin{equation}
2\lambda \rho'' r(b-r)+\lambda \rho'(rb'+3b-4r)
    +\rho(r^2+2\lambda b')-b'=0\,,\label{field3ai}
\end{equation}
\begin{equation}
4\lambda r\rho'(b-r)+2\lambda \rho(b-b'r)
   -rp_r(r^2+2\lambda b')-b=0\,,
\label{field3aii}
\end{equation}
\begin{eqnarray}
4\lambda r^2\rho'' (b-r)+2\lambda r\rho'(rb'+b-2r)
    \nonumber  \\
-2\lambda \rho(rb'+b)-2rp_t\left(r^2+2\lambda b'\right)+b-rb'=0\,. \label{field3aiii}
\end{eqnarray}
Note that the specific case of $\lambda=0$ reduces to the stress-energy distribution threading the wormhole in general relativity.

The violation of the NEC, i.e., $T_{\mu\nu}^{{\rm eff}}\,k^\mu k^\nu < 0$, in the context of the nonminimal curvature-matter coupling takes the following form
\begin{eqnarray}
\rho^{{\rm eff}}+p_r^{{\rm eff}} &=&\frac{1}{r^2}\Big[-2\lambda r^2 \rho''\left(1-\frac{b}{r}\right)
    \nonumber  \\
&&\hspace{-2.65cm}
+(\rho+p_r)\left(r^2+2\lambda b'\right)+\lambda \left(r\rho'+2\rho \right)\left(\frac{b'r-b}{r}\right)\Big]<0.
     \label{NECeff}
\end{eqnarray}
Analysed at the throat, taking into account the finiteness of the factor $\rho''$ at the throat, one has the following general condition $(\rho_0+p_{r0})\left(r_0^2+2\lambda b'_0 \right)
<\lambda\left(r_0\rho'_0+2\rho_0 \right)(1-b'_0)$.

Due to the nonlinearity of the equations, it is extremely difficult to obtain explicit exact solutions to the gravitational field equations. Nevertheless, the problem is mathematically well-defined in that one has three equations, with four functions, namely, the field equations (\ref{field3ai})-(\ref{field3aiii}), with four unknown functions of $r$, i.e., $\rho(r)$, $p_r(r)$, $p_t(r)$ and $b(r)$. It is possible to adopt different strategies to construct solutions with the properties and characteristics of wormholes. In particular, one may consider a meaningful equation of state, $p_r=p_r(\rho)$, or simply impose one of the unknown functions (see \cite{modgravity2} for more details).

\subsection{Wormholes in hybrid metric-Palatini gravity}

A new class of modified theories of gravity, consisting of the superposition of the metric Einstein-Hilbert Lagrangian with an $f(\cal R)$ term constructed \`{a} la Palatini was also proposed recently \cite{Harko:2011nh}. The dynamically equivalent scalar-tensor representation of the model was formulated, and it was shown that even if the scalar field is very light, the theory passes the Solar System observational constraints. Therefore, the model predicts the existence of a long-range scalar field that modifies the cosmological and galactic dynamics. An explicit model that passes the local tests and leads to cosmic acceleration was obtained, and cosmological solutions were further analysed \cite{Capozziello:2012ny}.

The action for the hybrid metric-Palatini gravity is provided by \cite{Harko:2011nh}
\begin{equation} \label{eq:S_hybrid}
S=\frac{1}{2\kappa^2}\int d^4 x \sqrt{-g} \left[ R + f({\cal R})\right] +\int d^4x\sqrt{-g}\;{\cal L}_m(g_{\mu\nu},\psi),
\end{equation}
where $\kappa^2\equiv 8\pi G$, $R$ is
the metric Einstein-Hilbert term, ${\cal R}  \equiv g^{\mu\nu}{\cal R}_{\mu\nu} $ is
the Palatini curvature, and ${\cal R}_{\mu\nu}$ is defined in terms of
an independent connection $\hat{\Gamma}^\alpha_{\mu\nu}$  as
\begin{equation}
{\cal R}_{\mu\nu} \equiv \hat{\Gamma}^\alpha_{\mu\nu ,\alpha} -
\hat{\Gamma}^\alpha_{\mu\alpha , \nu} +
\hat{\Gamma}^\alpha_{\alpha\lambda}\hat{\Gamma}^\lambda_{\mu\nu}
-\hat{\Gamma}^\alpha_{\mu\lambda}\hat{\Gamma}^\lambda_{\alpha\nu}\,.
\end{equation}

It was shown that the action (\ref{eq:S_hybrid}) can be expressed in the following
scalar-tensor representation (we refer the reader to \cite{Harko:2011nh} for details)
\begin{equation} \label{eq:S_scalar2}
S= \frac{1}{2\kappa^2}\int d^4 x \sqrt{-g} \left[ (1+\phi)R +\frac{3}{2\phi}\partial_\mu \phi \partial^\mu \phi
-V(\phi)\right] +S_m  \,,
\end{equation}
where $S_m$ is the matter action. It is important to note that this action differs fundamentally from the $w=-3/2$ Brans-Dicke theory in the coupling of the scalar to the curvature.

As in the examples above, the metric field equation can be written as an effective Einstein field equation, i.e.,  $G_{\mu\nu}= T^{\rm eff}_{\mu\nu}$, where for simplicity $\kappa^2=1$ and the effective stress-energy tensor is given by
\begin{eqnarray}
T^{\rm eff}_{\mu\nu}&=&\frac{1}{1+\phi} \Bigg\{ T_{\mu\nu}
 -  \Bigg[ \frac{1}{2}g_{\mu\nu}\left(V+2\Box\phi\right)+
     \nonumber \\
&& \nabla_\mu\nabla_\nu\phi-\frac{3}{2\phi}\partial_\mu \phi
\;\partial_\nu \phi + \frac{3}{4\phi}g_{\mu\nu}(\partial \phi)^2 \Bigg] \  \Bigg\}  \label{effSET} .
\end{eqnarray}

The scalar field is governed by the effective Klein-Gordon equation (we refer the reader to \cite{Harko:2011nh} for more details)
\begin{equation}\label{eq:evol-phi}
-\Box\phi+\frac{1}{2\phi}\partial_\mu \phi \partial^\mu
\phi+\frac{\phi[2V-(1+\phi)V_\phi]} {3}=\frac{\phi}{3}T\,.
\end{equation}
This relationship shows that, unlike in the Palatini ($w=-3/2$) case, the scalar field is dynamical.

In modified gravity, as considered above, it is the effective stress-energy tensor that violates the NEC at the throat, $T^{\rm eff}_{\mu\nu} k^\mu k^\nu |_{r_0} < 0$. The latter provides the following constraint in the present hybrid metric-Palatini gravitational theory
\begin{eqnarray}
T^{\rm eff}_{\mu\nu}k^\mu k^\nu |_{r_0} &=&\frac{1}{1+\phi} \Big\{ T_{\mu\nu} k^\mu k^\nu
 - \Big[k^\mu k^\nu \nabla_\mu\nabla_\nu\phi
     \nonumber \\
&& -\frac{3}{2\phi}k^\mu k^\nu\,\partial_\mu \phi
\;\partial_\nu \phi \Big] \  \Big\} \Big |_{r_0} < 0 \label{NECeffSET} .
\end{eqnarray}

Assuming that $1+\phi> 0$ and that standard matter satisfies the energy conditions and, in particular, the NEC, i.e, $T_{\mu\nu} k^\mu k^\nu \geq 0$,  one finds the generic constraint for hybrid metric-Palatini wormhole geometries
\begin{equation}
0\leq  T_{\mu\nu} k^\mu k^\nu |_{r_0} <  \Big[k^\mu k^\nu \nabla_\mu\nabla_\nu\phi
 - \frac{3}{2\phi}k^\mu k^\nu\,\partial_\mu \phi
\;\partial_\nu \phi \Big]  \Big|_{r_0}.  \label{NECeffSET2}
\end{equation}

Using the metric (\ref{metric}), the effective Einstein field equation provides the following gravitational field equations
\begin{eqnarray}
\rho(r) &=& \frac{b'}{r^2}(1+\phi) - \left(1-\frac{b}{r} \right)
\left[\phi'' -\frac{3(\phi')^2}{4\phi}  \right]
    \nonumber  \\
&&+\frac{\phi'}{2r}\left( b' + \frac{3b}{r}-4  \right) -\frac{V}{2} \,,
  \label{hybrid_rho}
  \\
p_r(r)&=& \left[-\frac{b}{r^3}+ \frac{2\Phi'}{r}\left(1-\frac{b}{r} \right)\right](1+\phi)
   \nonumber  \\
&& + \phi' \left(\Phi' + \frac{2}{r}
 + \frac{3 \phi'}{4\phi} \right)
\left( 1-\frac{b}{r} \right) + \frac{V}{2} \,, \\
p_t(r)&=&\Bigg[ \left( \Phi'' + (\Phi')^2 + \frac{\Phi'}{r}  \right)
\left(1-\frac{b}{r} \right)
   \nonumber   \\
&&+ \frac{b-b'r}{2r^3}\left(1+r\Phi' \right) \Bigg] (1+\phi)
  \nonumber  \\
&&+ \left[\phi''+\phi'\Phi' + \frac{3(\phi')^2}{4\phi} \right]
 \left( 1- \frac{b}{r} \right)
  \nonumber   \\
  &&
   +\frac{\phi'}{r}\left(1- \frac{b+rb'}{2r}  \right)  + \frac{V}{2} \,.
\end{eqnarray}

The effective Klein-Gordon equation (\ref{eq:evol-phi}) is given by
\begin{eqnarray}
\left[ \phi''+ \phi' \Phi' -\frac{(\phi')^2}{2\phi} + \frac{3\phi'}{2r}  \right] \left(1-\frac{b}{r} \right) 
   \nonumber \\
+ \frac{\phi'}{2r}(1+b') +  \frac{\phi}{3}\left[ 2V - (1+\phi)V_\phi \right]
  = \frac{\phi}{3} T \,.
        \label{modKGeq}
\end{eqnarray}

One may use several strategies in solving these field equations, namely, note that Eqs.~(\ref{hybrid_rho})-(\ref{modKGeq}) provide four independent equations, for seven unknown quantities, i.e. $\rho(r)$, $p_r(r)$, $p_t(r)$, $\Phi(r)$, $b(r)$, $\phi(r)$ and $V(r)$. Thus, the system of equations is under-determined. One may reduce the number of unknown functions by assuming suitable conditions. In fact, two specific examples were presented in \cite{Capozziello:2012hr}. In the first solution, the redshift function, the scalar field and the potential were specified. The solution found was not asymptotically flat and needs to be matched to a vacuum solution. In the second example, by adequately specifying the metric functions and choosing the scalar field, an asymptotically flat spacetime was found. We refer the reader to \cite{Capozziello:2012hr} for more details.

\section{Closed timelike curves and wormhole spacetimes}


One of the most fascinating aspects of wormholes is their apparent
ease in generating CTCs \cite{mty}. The basic idea is to create a time
shift between both mouths. This is done through the time dilation
effects in special relativity or in general relativity. More specifically, one
may consider the analogue of the twin paradox, in which the mouths
are in relative motion with respect to the other, or simply the case in
which one of the mouths is placed in a strong gravitational field \cite{Visser}.

\begin{figure*}[t]
\centering
  \includegraphics[width=2.4in]{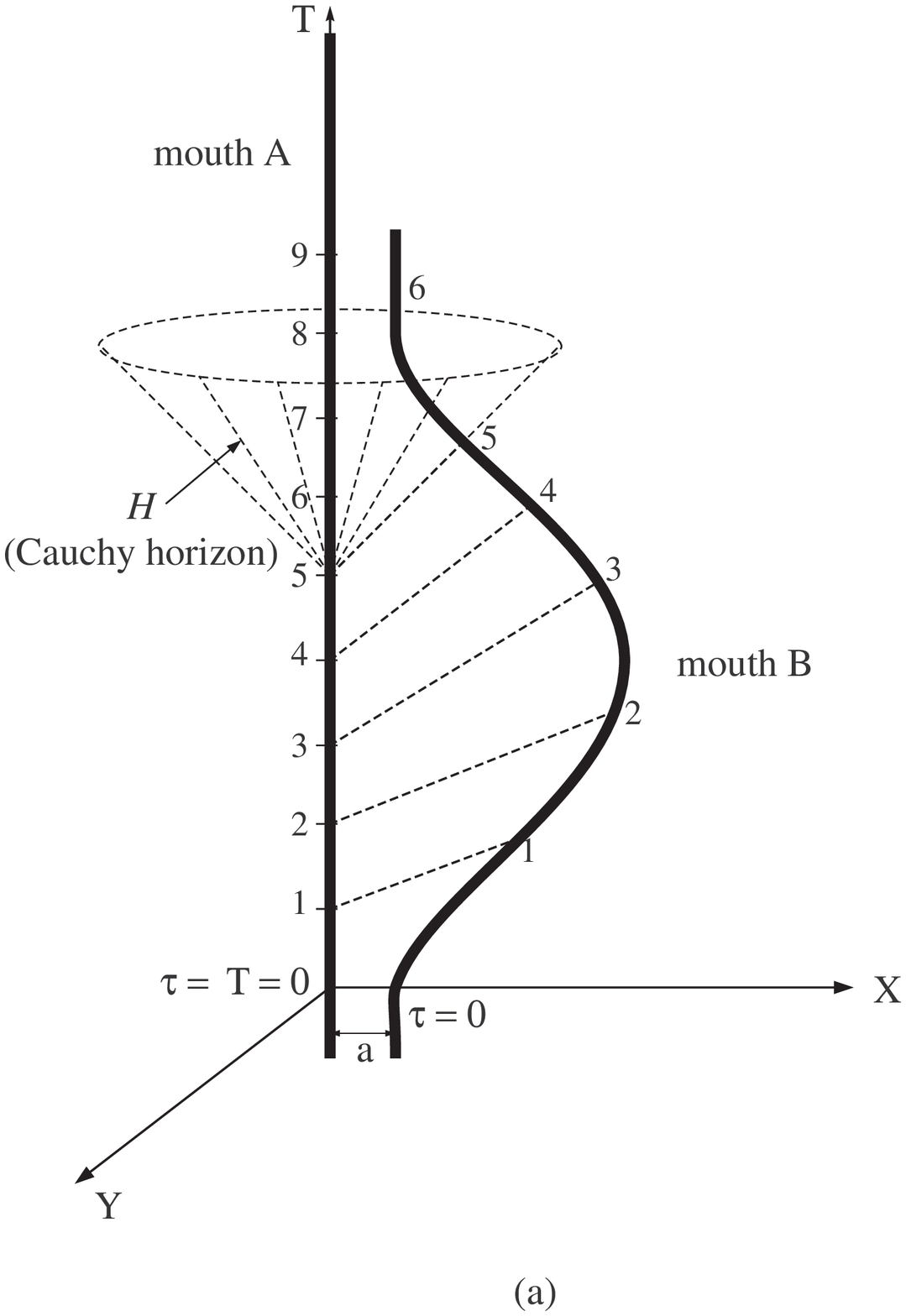}
  \hspace{0.6in}
  \includegraphics[width=2.2in]{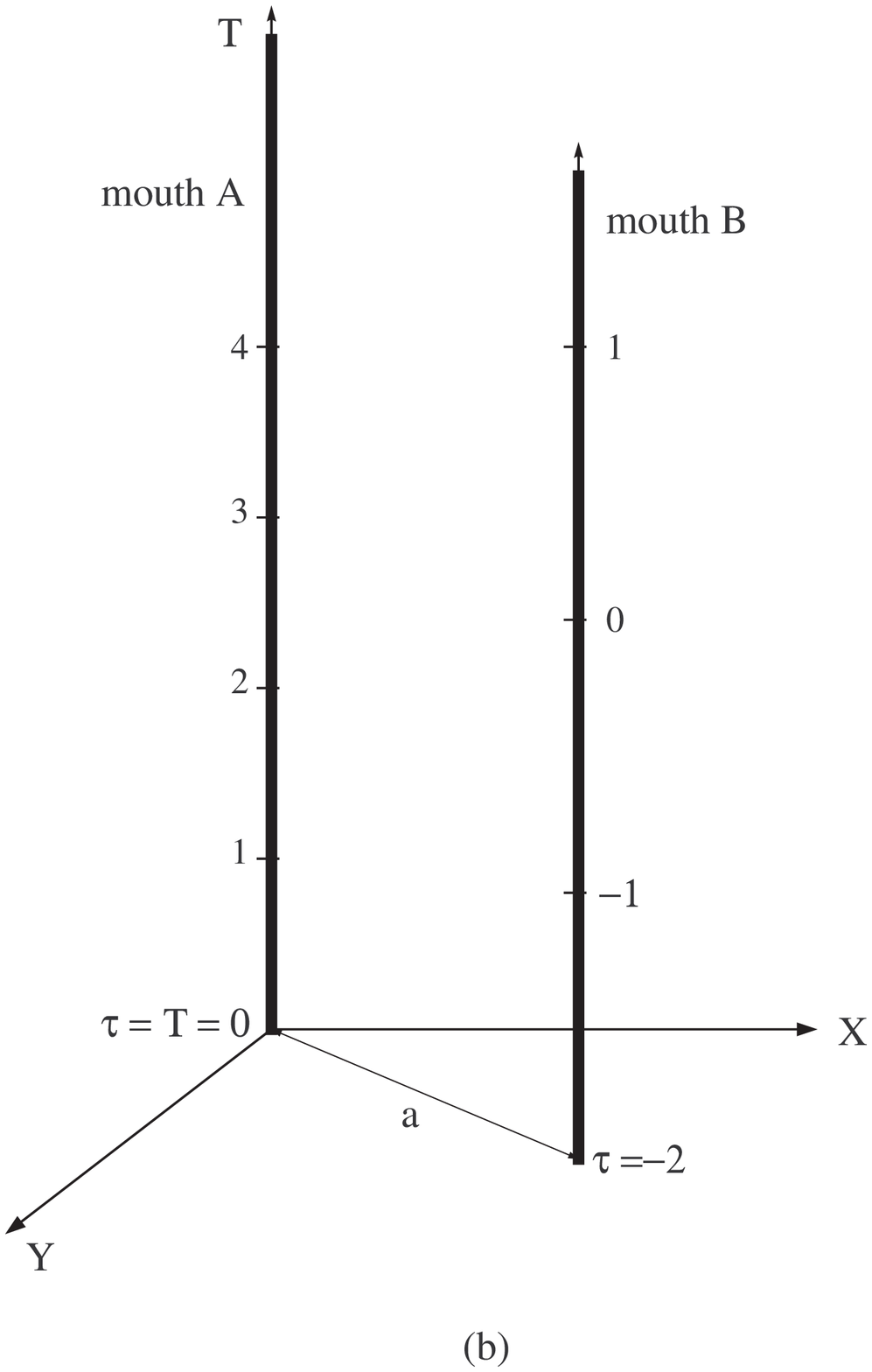}
  \caption[Wormhole spacetimes with closed timelike curves]
  {Depicted are two examples of wormhole spacetimes with CTCs. 
  The wormhole tunnels are arbitrarily short, and
  the two mouths are depicted as thick
  lines in the figure. Proper time $\tau$ at the wormhole throat is
  marked off, and identical values are the same event as seen
  through the wormhole handle. In Figure $(a)$, mouth $A$ remains at rest,
  while mouth $B$ accelerates from $A$ at a high velocity, then
  returns to its starting point at rest. A time shift is induced
  between both mouths, due to special relativistic time dilation effects. 
  The light cone-like hypersurface ${\it H}$ shown is
  a Cauchy horizon. Through every event to the future of ${\it H}$
  there exist CTCs, so that predictability breaks down. In Figure $(b)$, a time shift between both mouths
  is induced by placing mouth $B$ in strong gravitational field.
  }
  \label{fig:WH-time-machine}
\end{figure*}

As a specific example, consider the creation of a time shift using the twin paradox analogue. We assume that the wormhole mouths are in relative motion with respect to each other in external space, without significant changes to the internal geometry of the tunnel. For simplicity, consider that mouth $A$ is at rest in an inertial frame, while the other mouth $B$, initially at rest practically close by to $A$,
starts to move out with a high velocity, then returns to its starting point. Due to the Lorentz time contraction, the time interval between these two events, $\Delta T_B$, measured by a clock comoving with $B$ can be made to be significantly shorter than the time interval between the same two events, $\Delta T_A$, as measured by a clock resting at $A$. Thus, the clock that has moved has been slowed by the quantity $\Delta T_A-\Delta T_B$ relative to the standard inertial clock. Now, as the wormhole tunnel, connecting the mouths $A$ and $B$ remains practically unchanged, an observer comparing   clocks through the tunnels will measure an identical time, as the mouths are at rest with respect to one
another. However, by comparing the time of the clocks in external space, he will verify that their time shift is precisely $\Delta T_A-\Delta T_B$, as both mouths are in different reference frames. Consider an observer starting off from $A$ at an instant $T_0$, measured by the clock $A$. He travels to $B$ in external space and enters the tunnel from $B$. For simplicity, consider that the trip through the wormhole tunnel is instantaneous. He then exits from the wormhole mouth $A$ into external space at the instant $T_0-(\Delta T_A-\Delta T_B)$ as measured by a clock positioned at $A$. His arrival at $A$
precedes his departure, and the wormhole has been converted into a time machine. See Figure \ref{fig:WH-time-machine}.


The presence of CTCs apparently violate causality, opening Pandora's box and producing time travel paradoxes. The notion of causality is fundamental in the construction of physical theories, therefore time travel and it's associated paradoxes have to be treated with great caution. The paradoxes fall into two broad groups, namely the {\it consistency paradoxes} and the {\it causal loops}.

The consistency paradoxes occur whenever possibilities of changing events in the past arise. The grandfather paradox is the classical example, where a time traveller journeys into the past and murders his grandfather, before his father was born. The paradoxes associated to causal loops are related to
self-existing information or objects, trapped in spacetime. Much has been written on these two issues and on the possible remedies to the paradoxes, namely, the Principle of  Self-Consistency and the Chronology Protection Conjecture (see \cite{Lobo:2010sz} for a recent review).

One current of thought, led by Igor Novikov, is the Principle of Self-Consistency, which stipulates that events on a CTC are self-consistent, i.e., events influence one another along the curve in a cyclic and self-consistent way. Hawking's Chronology Protection Conjecture \cite{hawking} is a more conservative way of dealing with the paradoxes. Hawking notes the strong experimental evidence in favor of the conjecture from the fact that "we have not been invaded by hordes of tourists from the future". An analysis shows that as CTCs are about to be formed, the value of the renormalized expectation quantum stress-energy tensor diverges, so that the wormhole structure blows up. This conjecture permits the existence of traversable wormholes, but prohibits the appearance of CTCs.

\section{Conclusion}

In this paper, we have briefly explored the possibility that wormholes be supported by several modified theories of gravity, in particular, in $f(R)$ gravity, with a non-minimal curvature coupling and in the recently proposed hybrid metric-Palatini theory. In principle, matter threading the wormhole satisfies the energy conditions, and it is the higher order curvature derivative terms that support these nonstandard wormhole geometries, fundamentally different from their counterparts in general relativity.
We have also briefly considered the conversion of wormholes into time-machines, explored several of the time travel paradoxes and considered two possible remedies to these intriguing side-effects in wormhole physics.

%
%
%

\section*{Acknowledgements}
FSNL acknowledges financial support of the Funda\c{c}\~{a}o para a Ci\^{e}ncia e Tecnologia through the grants CERN/FP/123615/2011 and CERN/FP/123618/2011.

\end{document}